# Bridging the Analytics Divide: Retail Technology Diffusion in South Africa's Traditional Retail Sector

Shaun Moloi[1[0000-0003-2065-408X]], Adheesh Budree[2[0000-0002-7448-4453]] and Pitso Tsibolane[3[0000-0002-7888-1181]]

[1] University of Cape Town, South Africa mlxtum003@myuct.ac.za
[2] University of Cape Town, South Africa adheesh.budree@uct.ac.za
[3] University of Cape Town, South Africa pitso.tsibolane@uct.ac.za

**Abstract.** Retail analytics holds immense potential for enhancing decision-making, efficiency, and competitiveness in the retail sector. However, its adoption across traditional and hybrid retail industries in emerging markets remains inconsistent. This study examines the adoption of analytics technologies in South Africa's traditional retail sector, utilizing the Technology-Organization-Environment (TOE) framework and the Diffusion of Innovations (DOI) theory. Drawing on a qualitative multi-case study of fifteen executives *(n = 15)* spanning three retail organizations, the study identifies technological constraints (legacy systems, lack of IT infrastructure), organizational barriers (limited analytic capability, cultural resistance), and environmental factors (regulatory uncertainty, socio-economic inequality) as significant impediments to analytics adoption. Despite these challenges, findings reveal that phased analytics implementation, executive sponsorship, and alignment with core retail operations facilitate successful integration. The study provides a specialized framework for analytics diffusion in constrained environments and recommends strategies for policy and practice that align with the realities of developing economies.

## 1 Introduction

The adoption of analytics in retail has transformed the global marketplace, enabling firms to gain granular insights into customer behavior, optimize inventory, personalize marketing, and enhance strategic decision-making. In digitally mature economies, retail analytics has become a key driver of competitiveness, contributing to improved supply chain agility, revenue growth, and operational efficiency [1, 2]. Tools such as predictive modelling, real-time dashboards, and machine learning algorithms are now widely embedded in enterprise retail systems [3]. However, the picture is vastly different in many developing countries where the integration of analytics into traditional retail operations remains limited and fragmented.

In emerging markets like South Africa, retailers face a dual economy: on the one hand, large corporate chains such as Shoprite and Woolworths have begun incorporating analytics into business processes, while on the other, a wide base of traditional and mid-tier retailers continue to operate with outdated point-of-sale systems, manual inventory tracking, and limited access to reliable data infrastructure [8, 16]. This divergence, which the current study refers to as the analytics divide, highlights not only gaps in technology adoption but also deeper organizational, cultural, and regulatory barriers that constrain digital transformation.

Understanding this divide is important as retail in South Africa is not just a commercial activity but a crucial socio-economic engine. The sector contributes significantly to employment, consumer access, and informal trade networks, especially in peri-urban and township economies [16]. Yet, research on analytics adoption in these segments remains scarce, with most literature focusing on large multinationals in stable, digitally advanced environments [8]. Furthermore, existing adoption models do not adequately address the unique institutional constraints found in emerging contexts such as erratic electricity supply, low digital literacy, and evolving data privacy regulations like South Africa's Protection of Personal Information Act (POPIA) [5, 18].

To explore these dynamics, this study draws on two theoretical frameworks: the Technology-Organization-Environment (TOE) framework and the Diffusion of Innovations (DOI) theory. The TOE framework identifies three dimensions namely technological, organizational, and environmental that shape a firm’s readiness and ability to adopt new innovations [10]. Meanwhile, DOI theory focuses on the characteristics of the innovation itself (e.g., relative advantage, complexity, trialability) and the social system in which it diffuses [13]. Together, these frameworks provide a comprehensive lens through which to examine the layered and context-specific challenges faced by retailers seeking to integrate analytics into their business models.

The empirical core of the research is a multi-case study of three South African retailers representing different subsectors and levels of digital maturity. Through interviews, internal documents, and observational data, the study aims to answer the central research question: How do South African traditional and mid-tier retailers navigate the adoption and diffusion of analytics technologies, and what factors enable or hinder this process? In addressing this question, the study contributes to a growing body of work that seeks to make technology adoption models more inclusive of developing world realities.

This paper is structured as follows: Section 2 reviews the literature on analytics adoption, TOE/DOI theory, and the South African retail landscape. Section 3 details the research design and data collection process. Section 4 presents the findings categorized according to TOE dimensions. Section 5 interprets these findings using both TOE and DOI theories and proposes a practical diffusion model tailored to emerging market conditions. Section 6 concludes with a synthesis of implications for practitioners, policymakers, and scholars.

## 2 Literature Review

The objective of this section is to establish a theoretical foundation for understanding analytics adoption in the retail sector, particularly in emerging market contexts. The review proceeds in five parts: examining the role of analytics in retail transformation; highlighting the distinctive adoption challenges in emerging markets; presenting the TOE and DOI frameworks as conceptual lenses; and contextualizing these themes within the unique conditions of South African retail.

### 2.1 Retail Analytics and Digital Transformation

Retail analytics refers to the application of data science, machine learning, and business intelligence techniques to retail operations, with the goal of enhancing decision-making and performance. In mature economies, analytics has been a central feature of digital transformation strategies across the sector. Firms routinely deploy technologies that enable real-time inventory tracking, customer segmentation, demand forecasting, price optimization, and personalized marketing [1, 2]. These tools provide actionable insights by aggregating and analyzing data from diverse sources, including POS systems, online transactions, loyalty programs, and customer feedback platforms.

The significance of retail analytics extends beyond operational efficiency. As Wamba et al. [3] argue, analytics contributes to strategic agility by enabling firms to sense and respond to market changes faster than competitors. The integration of analytics into enterprise resource planning (ERP) and customer relationship management (CRM) systems reflects its maturation into a core business capability. Furthermore, analytics is increasingly deployed in omnichannel contexts, where seamless integration across physical and digital platforms is critical for competitive positioning [4]. In these environments, analytics has become not merely a support tool but a strategic asset capable of shaping long-term value creation.

Despite the demonstrated benefits, the successful deployment of analytics is contingent on several organizational enablers, including strong IT infrastructure, skilled human capital, and supportive leadership [5]. Without these, firms may struggle to extract value from analytics investments, and their transformation efforts may stall or regress. Most of the existing literature assumes the presence of these enabling conditions, focusing on case studies from the United States, Europe, and East Asia, where such capabilities are more widespread.

### 2.2 Analytics in Emerging Market Contexts

In contrast to the analytics-rich environments of developed markets, emerging economies face a set of structural and institutional barriers that complicate the adoption of data-driven technologies. These barriers include inadequate infrastructure, fragmented data systems, limited internet access, unreliable electricity supply, and low levels of digital literacy among workers [6, 7]. Many retailers in these contexts rely on legacy systems or manual processes, making the leap to analytics-intensive operations particularly challenging.

Empirical studies suggest that firms in emerging markets also contend with weak regulatory clarity, economic instability, and high operational risk, all of which contribute to a cautious and incremental approach to innovation [8]. The return on investment for analytics is often difficult to quantify in such environments, especially when basic digitization has yet to be achieved. The result is a paradox: while analytics offers significant potential to streamline operations and reach underserved customer segments, the capability to implement it is often absent or underdeveloped.

Nevertheless, there are documented cases of analytics adoption in resource-constrained settings. Ghosh et al. [9], for example, examine African retailers that have used cloud-based analytics tools to overcome hardware limitations and access real-time data insights. Such cases underscore the importance of context-sensitive implementation strategies, including phased rollouts, external partnerships, and embedded training. These adaptations help mitigate risks and enable firms to gradually build analytics maturity. However, the broader literature still lacks robust, theory-informed studies that systematically address the adoption of analytics in emerging retail sectors, a gap this study aims to fill.

### 2.3 Technology–Organization–Environment (TOE) Framework

The Technology–Organization–Environment (TOE) framework, developed by Tornatzky and Fleischer [10], offers a holistic model for analyzing the factors that influence organizational adoption of innovations. The framework is composed of three interrelated dimensions. The technological context includes characteristics of the existing and emerging technologies available to the firm, such as compatibility, complexity, and relative advantage. The organizational context comprises internal attributes such as firm size, structure, managerial capabilities, and available resources. Finally, the environmental context involves external pressures and influences, including competition, regulatory frameworks, and technological infrastructure.

One of the strengths of the TOE framework is its flexibility; it can be applied across industries and technology types, making it particularly useful for analyzing emerging digital tools such as analytics platforms. It has been widely employed in studies of enterprise system adoption, e-commerce integration, and cloud computing, particularly in developing countries [11]. For instance, Alshamaila et al. [11] use the TOE framework to examine cloud adoption by SMEs in the UK, finding that organizational readiness and environmental uncertainty played significant roles in shaping adoption outcomes.

In the context of this study, the TOE framework provides a structured approach to investigating the conditions that support or hinder analytics adoption. By considering technological constraints (e.g., legacy systems), organizational barriers (e.g., limited skills), and environmental uncertainties (e.g., regulatory ambiguity), the framework captures the complexity of innovation processes in under-resourced contexts. However, while TOE is valuable in identifying structural and contextual variables, it does not fully capture the perceptual and behavioral aspects of innovation diffusion, which are addressed more explicitly in the DOI theory.

### 2.4 Diffusion of Innovation (DOI) Theory

Rogers' Diffusion of Innovation (DOI) theory [13] complements the TOE framework by focusing on how innovations spread within a social system. DOI posits that adoption is influenced by five key characteristics of the innovation: relative advantage (the perceived benefit over current practice), compatibility (alignment with existing values and needs), complexity (ease of use), trialability (the ability to experiment), and observability (the visibility of results). These attributes shape user perceptions and behaviors, ultimately determining the pace and extent of diffusion.

DOI also introduces a typology of adopters namely innovators, early adopters, early majority, late majority, and laggards to explain variation in adoption timelines. In organizational contexts, this typology is useful for understanding the role of leadership, change agents, and internal champions in driving technology uptake. For example, studies in retail have shown that analytics initiatives often begin with visionary leaders who can articulate the business case and mobilize resources, before being institutionalized across departments [14].

Despite its contributions, DOI has limitations when applied to contemporary technology adoption contexts. Recent empirical evidence challenges the universality of Rogers' diffusion patterns. Comin and Mestieri analyzed technology diffusion across 166 countries and found that traditional S-shaped diffusion curves fail to fit 53% of technology-country cases when incorporating intensity of use [15]. They demonstrate that "traditional diffusion measures have two important drawbacks: they do not capture the intensity with which each adopter uses the technology" and often miss the complexity of adoption in resource-constrained environments. This critique is particularly relevant in emerging markets where structural barriers, such as infrastructure deficits, regulatory uncertainty, and resource limitations, significantly shape adoption trajectories in ways that individual-level perception models cannot fully capture. As such, combining DOI with TOE provides a more complete analytical lens, balancing structural determinants with cognitive and behavioral insights [15].

### 2.5 South African Retail and the Analytics Divide

South Africa's retail sector is a complex and stratified ecosystem that includes large corporate chains, informal spaza shops, township convenience stores, and mid-sized regional players. This diversity creates significant variation in technology use and analytics capability. While major retail chains like Pick n Pay and Woolworths have made investments in digital transformation and customer analytics, smaller and mid-tier retailers often operate with limited infrastructure and human resource constraints [16, 17]. This internal divide mirrors broader socio-economic inequalities in South African society and creates what may be termed an "analytics divide", a gap not only between countries but within the national economy itself.

Regulatory developments such as the Protection of Personal Information Act (POPIA) have added new dimensions to this divide. While the legislation aligns with international data privacy standards, its implementation has created confusion among businesses, particularly SMEs, that lack legal or compliance teams. Many retailers are unsure how to collect, store, and process customer data in a compliant manner, leading to hesitancy in deploying analytics tools that rely on personal information [18]. At the same time, consumers in South Africa are becoming more digitally engaged, particularly younger demographics in urban areas who expect personalized offers, mobile integration, and seamless shopping experiences.

Despite these challenges, there are emerging signs of progress. Local tech startups, cloud-based service providers, and university-industry collaborations are beginning to offer analytics solutions tailored to the South African context. However, empirical research remains limited. As Mlitwa and Nonyana [8] note, few studies have examined how traditional retailers in developing contexts make decisions about adopting complex technologies like analytics, or how these decisions are shaped by local realities. This study responds to that gap by offering a grounded, theory-informed exploration of analytics adoption in a sector that is critical to South Africa's economic and social fabric.

## 3 Methods

This study employed a qualitative case study methodology situated within an interpretive research paradigm. The aim was to explore how traditional and mid-tier retailers in South Africa experience, interpret, and respond to the challenges of analytics adoption. Given the socio-technical complexity of the phenomenon under investigation, involving organizational routines, technological systems, and contextual constraints, a qualitative approach was deemed most appropriate for generating rich, contextual insights. The methodology was designed to capture multiple perspectives across different organizational levels and to trace how analytics practices evolve in real-world settings. The following sections detail the research design, case selection rationale, data collection techniques, ethical considerations, analytical procedures, and the steps taken to ensure research validity and trustworthiness.

### 3.1 Research Design

The study was anchored in an interpretive research paradigm, which views reality as socially constructed through interactions and meanings developed through experience [20]. As Myers and Klein emphasize, interpretive research requires an explicit acknowledgment of how the researcher's philosophical stance shapes both data collection and analysis [32]. Rather than seeking universal laws or statistical generalizability, the goal is to understand how phenomena are perceived and enacted by participants within their organizational and cultural contexts. This approach is particularly suited to the study of analytics adoption, which is influenced not only by technical capability but also by subjective interpretations of value, complexity, and risk.

Following Walsham's guidance on interpretive case studies, we recognize that our understanding of analytics adoption is mediated through participants' lived experiences and our own interpretive lens [33]. As Greenhalgh et al. further argue, this perspective is particularly relevant in emerging markets where different groups "have different types of social networks” and “use different language and metaphors for diffusion, dissemination, and implementation” [34].

In line with this philosophical orientation, we adopted a multiple-case study design, consistent with the methodological guidance provided by Yin [21]. This design enables an in-depth exploration of a contemporary phenomenon within its real-world context using multiple sources of evidence. Each case serves as a discrete unit of analysis, allowing for detailed within-case analysis followed by cross-case synthesis. This approach enhances analytic generalizability through replication logic, as patterns confirmed or disconfirmed across different organizational settings help reveal broader trends and causal mechanisms.

This interpretive commitment shaped all stages of the research process, the data collection emphasized participant narratives over standardized metrics; our analysis focused on meaning-making rather than variable testing; and our theoretical development prioritized contextualized, rather than universal, models of technology adoption.

### 3.2 Case Selection

Purposive sampling was used to identify three cases that would yield deep, varied, and information-rich insights into analytics adoption in the South African retail sector. The inclusion criteria were designed to ensure theoretical relevance and practical diversity. First, each organization had to fall within the traditional or mid-tier segments of retail, where analytics diffusion is typically uneven. Second, organizations were required to have either initiated analytics-related projects or expressed strategic interest in such initiatives. Finally, the researcher had to gain access to multiple stakeholders across different organizational levels to obtain a holistic view of the phenomenon.

The three selected retailers operated in different product categories namely FMCG, clothing, and general merchandise, and represented varying degrees of digital maturity. Retailer A was a regional supermarket chain using legacy infrastructure, Retailer B had recently launched a pilot analytics tool in head office, and Retailer C was the most advanced, with an in-house digital team and partial analytics implementation. The sampling strategy was guided by theoretical sampling principles aimed at maximizing variance and comparative insight across settings [22].

### 3.3 Data Collection

Data were collected over a six-month period using a multi-method approach. The primary method was semi-structured interviews, which allowed for flexibility in probing different dimensions of analytics adoption while maintaining consistency across key themes. A total of 15 interviews were conducted, with participants ranging from executive leadership (e.g., CIOs, heads of operations) to middle managers (e.g., store managers, IT officers). Each interview lasted between 45 and 90 minutes, was audio recorded with consent, and subsequently transcribed verbatim.

Supplementary data were collected from organizational documents, such as internal analytics reports, digital strategy roadmaps, training materials, and compliance statements. These documents enriched the primary data by offering institutional context and helping validate claims made in interviews. Site visits were conducted in five locations, where operational practices were observed, and informal conversations with staff were recorded as field notes. This triangulation of data sources specifically interviews, documents, and observation enhanced the robustness and credibility of the study, as recommended in case study methodology [24].

Interview questions were guided by constructs from the Technology–Organization–Environment (TOE) framework and the Diffusion of Innovations (DOI) theory. The guides explored issues such as system complexity, organizational alignment, training and literacy, data governance, and perceived regulatory barriers. This theory-informed design ensured that data collection was systematic while still allowing room for emergent themes to surface.

### 3.4 Ethical Considerations

The study received formal ethics approval from the University of Cape Town's Faculty of Commerce Research Ethics Committee. All participants were provided with informed consent forms detailing the purpose of the study, their rights, and data handling procedures. Anonymity and confidentiality were strictly maintained, with pseudonyms used for both individuals and organizations (Retailers A, B, and C). Recordings and transcripts were securely stored on encrypted devices and accessible only to the research team.

Ethical fieldwork practices were guided by the principles of transparency, voluntary participation, and respect for the autonomy and well-being of participants. Given the sensitive nature of some discussions, particularly regarding POPIA compliance and internal data weaknesses the researcher maintained open communication with gatekeepers and offered participants the opportunity to review and clarify their statements prior to inclusion in the report. These practices align with established ethical protocols in qualitative information systems research [25].

### 3.5 Data Analysis

Thematic analysis was conducted using Braun and Clarke's six-step framework: data familiarization, initial coding, theme development, review, definition, and final write-up [26]. The process began with a close reading of transcripts and field notes, during which preliminary codes were assigned. Coding was both deductive based on TOE and DOI concepts and inductive, allowing new themes to emerge from the data. For example, participant concerns about job security and the perceived opacity of analytics tools were inductively coded under emerging themes like "automation anxiety" and "dashboard fatigue."

NVivo 12 was used for data management and coding. Each case was analyzed individually to surface within-case dynamics before conducting cross-case analysis. Comparative analysis allowed the researcher to identify recurring patterns, anomalies, and causal relationships across the three settings. This dual-stage process of analysis ensured theoretical saturation and enhanced analytic rigor [27]. Memo-ing was employed throughout the coding process to track interpretive decisions and reflexive insights.

Themes were continuously refined through iteration and triangulated with document and observational data. Key constructs such as "data literacy gaps," "legacy system constraints," and "regulatory uncertainty" were assessed across cases for convergence. Where differences existed, these were explored as explanatory outliers, deepening theory development.

### 3.6 Trustworthiness and Validity

To ensure the trustworthiness of the findings, the study followed Lincoln and Guba's four criteria: credibility, dependability, confirmability, and transferability [28].

**Credibility.** This was achieved through triangulation of multiple data sources and prolonged engagement in the field. Member checking was also employed, with several participants reviewing thematic summaries to confirm the accuracy of interpretations.

**Dependability.** This was established through careful documentation of research procedures and analytical steps, creating an audit trail that enhances transparency.

**Confirmability.** This was supported by maintaining a reflexive journal in which the researcher recorded decisions, impressions, and potential biases throughout the study. This reflexivity helped minimize subjective interference and bolstered the neutrality of the findings [29].

**Transferability.** While transferability in qualitative research is context-bound, this study supports it by providing thick descriptions of each case's environment, structure, and practices. These details enable readers to assess the relevance of the findings to similar contexts or settings. Overall, the methodological approach adheres to established standards for qualitative research in information systems, enhancing the rigor, credibility, and relevance of the work [30], [31].

## 4 Findings

This section presents an in-depth analysis of findings from the three case study retailers, detailing how analytics adoption is shaped by technological, organizational, and environmental dynamics. The findings are organized under key themes that emerged during data analysis and aligned with constructs from the Technology–Organization–Environment (TOE) framework and the Diffusion of Innovations (DOI) theory. Each subsection reflects cross-case patterns as well as distinctive contextual nuances observed during fieldwork.

### 4.1 Technological Context

Across all three retailers, the technological environment was characterized by a legacy infrastructure landscape and fragmented data ecosystems, which collectively undermined analytics integration. Retailer A operated on a decades-old ERP system with no real-time data interface, relying instead on weekly batch reports and ad hoc spreadsheet manipulation by individual departments. Participants in both operations and IT described this environment as "too rigid to experiment," reflecting a common sentiment that analytics required modern architectures to function effectively.

Retailer B had begun modernizing its systems but had not completed data consolidation across business units. Data silos persisted between merchandising, supply chain, and finance, which impeded the use of analytics for enterprise-level decision-making. Interviewees described how the lack of data interoperability produced confusion, duplication of effort, and “shadow analytics” where individual users developed their own disconnected dashboards, often in Excel or Power BI, without IT oversight or quality assurance.

Retailer C had the most developed technology stack, including a centralized data warehouse and basic self-service analytics tools available to managers. However, even in this comparatively advanced setting, limitations in system integration and data governance were evident. The investigation revealed that these inconsistencies often emerged from legitimate operational needs rather than mere oversight. For instance, the marketing department defined 'active customer' as 'any purchase within 30 days' to optimize campaign targeting, while finance defined it as 'account balance greater than zero' for credit risk assessment. Each definition served a specific, context-appropriate purpose. This reflects broader findings in analytics adoption literature, where fragmented digital infrastructure and definitions hinders value realization despite technical investment [4, 5].

Interviewees across all cases noted that analytics tools themselves were often "not intuitive" or "overly technical." The cognitive load associated with using tools like SQL-based dashboards or low-code predictive platforms was especially daunting for frontline users. This finding echoes prior research on the importance of technological usability in adoption success [13].

## 4.2 Organizational Context

Organizational capacity emerged as a pivotal influence on analytics readiness. In all three cases, there was a clear disparity between top-level strategic enthusiasm for analytics and actual operational enablement. Retailer A’s executives spoke of a "data-driven culture" as a strategic goal, yet no formal analytics governance, training programs, or performance incentives were in place. Store-level employees had minimal understanding of analytics beyond basic report interpretation, and many relied on intuition or precedent in decision-making.

At Retailer B, an internal digital transformation working group had been established, but interviewees described it as "isolated" from the operational core. Pilot analytics projects were often short-lived, as business units lacked the time, personnel, and change management support to institutionalize findings. This mirrors the implementation gaps observed in previous studies of mid-tier digital initiatives in resource-constrained firms [12, 15].

Retailer C’s analytics journey was further along, supported by a dedicated in-house data science team. However, the centralization of analytics expertise created a bottleneck. Most analytics queries had to be submitted as tickets, resulting in slow turnaround and disengagement from business units. Managers described the analytics team as "too technical and far removed" from the day-to-day context, leading to misalignment between analytical outputs and business needs. These dynamics reinforce the literature on organizational alignment as a success factor in analytics adoption [6, 24].

Skills gaps were pervasive. In all cases, interviewees described challenges with data literacy and a lack of internal capacity to interpret or question analytic outputs. While training programs existed in Retailer C, they were not mandatory or systematically aligned with evolving analytics capabilities. As a result, many staff continued to rely on reports and insights produced by others without interrogating assumptions or verifying data accuracy, a finding consistent with studies on analytics readiness in developing markets [7, 14].

### 4.3 Environmental Context

External pressures, particularly regulatory and competitive dynamics, significantly influenced analytics trajectories across the cases. All retailers cited the Protection of Personal Information Act (POPIA) as both a catalyst and constraint. In Retailer A, POPIA was interpreted as a legal risk that inhibited experimentation with customer-level data. Executives expressed caution around implementing personalization strategies due to concerns over data handling compliance and reputational risk. This "compliance anxiety" curtailed initiatives that would otherwise require more granular data analysis.

Retailer B responded by instituting a POPIA task force but struggled to embed data governance principles across business units. Interviewees described compliance as "someone else's job," indicating a low level of internalization. Conversely, Retailer C viewed regulation as an opportunity to improve data governance and build customer trust, with executives noting that formalizing data stewardship had improved internal data hygiene and stakeholder confidence. These responses align with broader insights that institutional environments shape both the pace and form of innovation diffusion [11, 17].

Market pressure also shaped analytics motivations. Retailer A, facing competition from digitally advanced chains, framed analytics as a potential "equalizer" but lacked a coherent execution roadmap. Retailer B used analytics for pricing simulations in pilot stores, but results were not scaled. Retailer C applied analytics in demand forecasting and customer segmentation but struggled to measure impact systematically. These findings align with evidence that perceived competitive urgency does not automatically translate into internal alignment or follow-through [6].

Finally, analytics vendors and consultants played a gatekeeping role. All three retailers had engaged with third-party providers at various stages, but experiences were mixed. Some vendors provided turnkey solutions without organizational capacity-building, while others lacked contextual understanding. One CIO described analytics consulting as "high on hype, low on handover," reflecting frustrations with externally driven adoption efforts that failed to embed internally. This tension reinforces the need for absorptive capacity, not just tool acquisition but as a foundation for analytics effectiveness [25, 26].

# 5 Discussion

The findings reveal complex interactions between technological, organizational, and environmental factors influencing analytics adoption in South African retailers. Applying the Technology–Organization–Environment (TOE) framework and Diffusion of Innovations (DOI) theory provides contextualized insights into structural and perceptual adoption determinants, particularly in resource-constrained environments.

## 5.1 Applying the TOE Framework

Technological factors demonstrated that system complexity, data fragmentation, and limited interoperability significantly inhibited adoption. Legacy infrastructure in Retailers A and B constrained real-time analytics deployment, while Retailer C's centralized data warehouse suffered from inconsistent data definitions and weak platform integration. These findings align with prior research identifying compatibility and system integration as critical to technological readiness [5, 10].

Organizational factors revealed systemic deficiencies in data literacy, change management, and cross-functional collaboration. Executive rhetoric about becoming "data-driven" lacked supporting investments in training, incentives, or analytics governance. Absent analytics stewardship and internal champions undermined strategy operationalization, supporting the TOE proposition that employee skills, managerial support, and internal communication strongly condition innovation assimilation [2, 12].

Environmental factors included regulatory pressure, market competitiveness, and vendor influence. POPIA compliance created hesitation around customer data use in Retailer A, while Retailer C framed compliance as a data governance opportunity. This disparity illustrates variable institutional interpretation patterns observed in emerging markets [11, 17]. Vendor involvement enabled initial analytics exposure but failed to embed internal capabilities, creating external dependency without absorptive capacity [25].

## 5.2 Applying DOI Theory

The DOI theory reveals how innovation perceptions affected adoption. Perceived complexity emerged prominently—managers described platforms as "intimidating" and "not user-friendly," inhibiting experimentation. According to DOI, complex innovations face adoption resistance, especially with weak training structures [1].

Relative advantage perception varied significantly. Retailer C managers associated analytics with improved forecasting, while Retailers A and B's lack of visible returns from prior initiatives created skepticism. Limited demonstrable benefits reduced perceived relative advantage, consistent with literature on low-maturity digital environments [6, 13].

Trialability and observability were constrained across cases. Siloed, short-lived, poorly documented pilots prevented benefit observation or success replication, undermining social diffusion. DOI suggests innovations with visible, trialable benefits spread more rapidly [1, 16].

Compatibility created adoption gaps where executive enthusiasm wasn't mirrored by grassroots engagement, as analytics disrupted established routines despite aligning with strategic aspirations [14].

### 5.3 Toward a Contextualized Analytics Adoption Model

Drawing from TOE and DOI lenses, this study proposes a contextualized interpretation of analytics readiness in resource-constrained retail environments. Technological readiness requires interoperability, governance, and usability beyond mere tool presence. Organizational readiness demands absorptive capacity to translate analytics into action beyond formal strategy. Environmental readiness involves navigating regulatory constraints, institutional ambiguity, vendor relationships, and socio-political considerations unique to emerging markets.

Analytics adoption represents an iterative, interpretive journey requiring alignment across technical systems, human capabilities, organizational routines, and institutional conditions. This interpretation challenges universalist maturity models, calling for grounded approaches emphasizing adaptation, co-design, and social learning over global best practice replication.

### 5.4 Analytics Diffusion Model

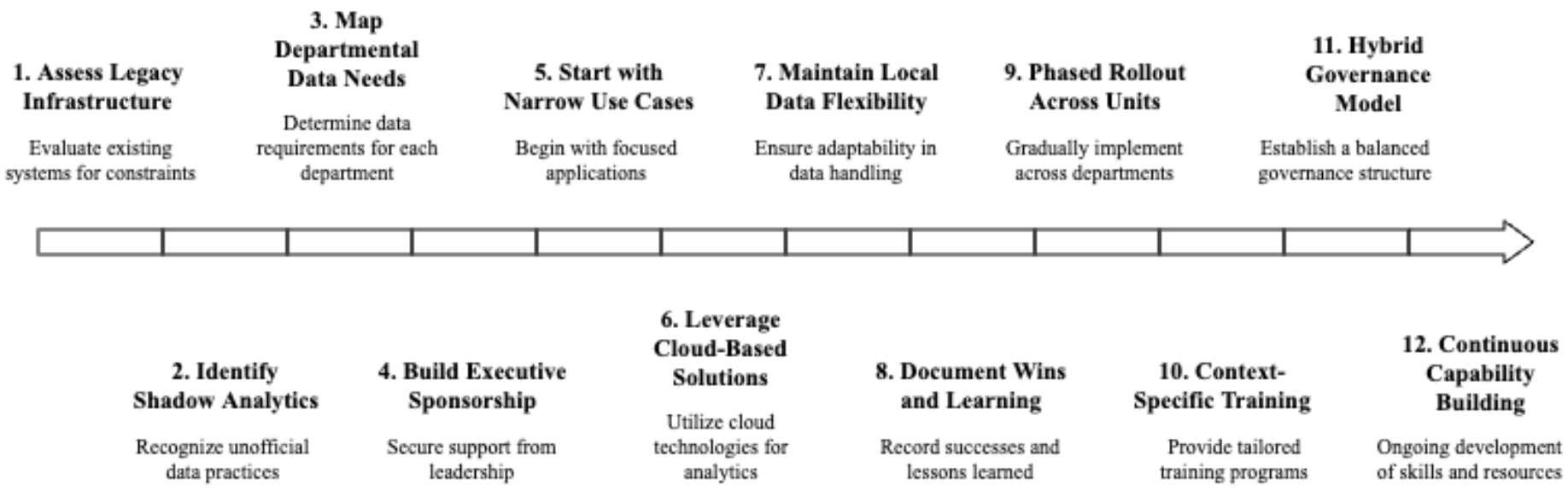

**Fig. 1.** Analytics Diffusion Model

Empirical findings reveal that successful analytics adoption in resource-constrained environments follows distinct patterns differing from traditional diffusion models. Rather than linear progression or symmetric adoption curves, this study proposes a three-stage model reflecting iterative, adaptive technology adoption under simultaneous technological, organizational, and environmental constraints.

The model comprises twelve key activities across three stages. Activities 1-4 constitute **Diagnostic Alignment**, preventing premature implementation failures. Activities 5-8 form **Strategic Piloting**, emphasizing controlled experimentation over comprehensive transformation. Activities 9-12 represent **Adaptive Scaling**, recognizing that expansion requires continuous adaptation rather than simple replication.

**Stage 1: Diagnostic Alignment (Activities 1-4)** focuses on understanding organizational starting positions through assessing legacy infrastructure constraints, identifying "shadow analytics" practices revealing latent demand, mapping departmental data needs, preventing standardization rendering metrics "useless," and building executive sponsorship securing genuine resource commitment.

**Stage 2: Strategic Piloting (Activities 5-8)** demonstrates that focused initiatives succeed where comprehensive transformations fail. This includes starting with narrow use cases, leveraging cloud-based solutions, formalizing existing tools, maintaining local data flexibility, preserving operational relevance, and documenting wins, addressing isolated pilot problems.

**Stage 3: Adaptive Scaling (Activities 9-12)** acknowledges that scaling requires ongoing adaptation through phased rollout based on readiness assessments, context-specific training addressing varied user perceptions, hybrid governance balancing standardization with flexibility, and continuous capability building overcoming acceptance of uninterrogated assumptions.

This model extends traditional diffusion theory by acknowledging that emerging market innovations require careful orchestration to overcome structural barriers, providing practitioners with realistic transformation roadmaps.

# 6 Conclusion

This study explored analytics adoption in South African traditional and mid-tier retail through TOE and DOI theoretical lenses. The research reveals complex, fragmented adoption landscapes shaped by technical constraints and cultural, institutional, and infrastructural realities.

**Technology:** Prominent barriers included outdated point-of-sale systems, incompatible data formats, and absent centralized infrastructure, creating real-time insight and predictive tool implementation challenges. Platform complexity perceptions compounded adoption friction, particularly with low data literacy.

**Organization:** Leadership support emerged as critical as retailers with executive analytics champions achieved coherent strategies and staff buy-in, while those treating analytics as peripheral IT functions encountered resistance. Data-driven culture presence significantly influenced outcomes, with employees struggling to interpret metrics or mistrusting outputs, which threatened established practices.

**Environment:** POPIA created dual impacts as initial uncertainty suspended promising pilots due to legal ambiguity, while simultaneously catalyzing improved data governance. Market competition from digitally savvy consumers spurred investment, though adoption remained uneven between urban and rural operations.

.

## 6.1 Implications

**Implication 1.** Analytics adoption requires phased, adaptive processes rather than comprehensive overhauls. Retailers beginning with strategically aligned pilots better demonstrated value and scaled initiatives, supported by contrasting experiences where Retailer B's comprehensive transformation failed due to operational isolation [Sections 4.1-4.3].

**Implication 2.** Internal capacity building is essential—analytics effectiveness depends on user capabilities. Contextualized data literacy training must accompany technology deployment, evidenced by Retailer C's technical-operational misalignment and persistent reliance on unverified reports [Section 4.2].

**Implication 3.** Cloud-based solutions effectively overcome infrastructure gaps through scalability and low entry costs, addressing documented infrastructure challenges and emerging "shadow analytics" trends [Section 4.1].

**Implication 4.** Hybrid data governance, balancing standardization with local flexibility, addresses legitimate operational variations while maintaining core consistency, as demonstrated by varying POPIA responses and departmental metric needs [Sections 4.1, 4.3].

### 6.2 Theoretical Contributions

This study demonstrates the utility of the TOE and DOI framework for understanding the adoption of holistic emerging market analytics. TOE provides structural insights while DOI enriches analysis through psychological and perceptual dimensions. The three-stage diffusion model (diagnostic alignment, strategic piloting, adaptive scaling) reflects constrained organization realities, offering blueprints for incremental rather than disruptive transformation.

### 6.3 Limitations

Qualitative, case-based design limits statistical generalizability but offers analytically generalizable insights for theory building and comparable settings. Future research could employ longitudinal tracking, cross-sector comparisons, or quantitative validation of TOE and DOI construct relationships.

Analytics diffusion in emerging market retail depends on leadership, cultural transformation, strategic alignment, and regulatory clarity beyond technological availability. Even under infrastructural and institutional constraints, thoughtful, incremental, inclusive analytics adoption remains achievable, contributing to the understanding of global South digital transformation and providing practical implementation roadmaps.

## References

1. Davenport, T.H., & Harris, J.G. (2007). Competing on Analytics: The New Science of Winning. Harvard Business Press.
2. Waller, M.A., & Fawcett, S.E. (2013). Data science, predictive analytics, and big data: A revolution that will transform supply chain design and management. Journal of Business Logistics, 34(2), 77–84.

3. Wamba, S.F., Akter, S., Edwards, A., Chopin, G., & Gnanzou, D. (2015). How 'big data' can make big impact: Findings from a systematic review and a longitudinal case study. International Journal of Production Economics, 165, 234–246.
4. Oliveira, T., & Martins, M.F. (2011). Literature review of information technology adoption models at firm level. Electronic Journal of Information Systems Evaluation, 14(1), 110–121.
5. Kshetri, N. (2014). Big data's impact on privacy, security and consumer welfare. Telecommunications Policy, 38(11), 1134–1145.
6. Chen, H., Chiang, R.H.L., & Storey, V.C. (2012). Business intelligence and analytics: From big data to big impact. MIS Quarterly, 36(4), 1165–1188.
7. World Bank. (2020). Doing Business 2020: Comparing Business Regulation in 190 Economies. Washington, DC.
8. Mlitwa, N., & Nonyana, D. (2021). Understanding digital readiness in South African retail. South African Journal of Information Management, 23(1), a1351.
9. Ghosh, R., Scott, J., & Thakur, R. (2022). Adoption of analytics in African retail: Challenges and opportunities. International Journal of Information Management, 66, 102603.
10. Tornatzky, L.G., & Fleischer, M. (1990). The Processes of Technological Innovation. Lexington Books.
11. Alshamaila, Y., Papagiannidis, S., & Li, F. (2013). Cloud computing adoption by SMEs in the north east of England. Journal of Enterprise Information Management, 26(3), 250–275.
12. Baker, J. (2012). The technology–organization–environment framework. In Information systems theory (pp. 231–245). Springer.
13. Rogers, E.M. (2003). Diffusion of Innovations (5th ed.). Free Press.
14. Karahanna, E., Straub, D.W., & Chervany, N.L. (1999). Information technology adoption across time: A cross-sectional comparison of pre-adoption and post-adoption beliefs. MIS Quarterly, 23(2), 183–213.
15. Comin, D., & Mestieri, M. (2014). Technology diffusion: Measurement, causes, and consequences. In Handbook of economic growth (Vol. 2, pp. 565-622). Elsevier.
16. Ligthelm, A.A. (2008). The impact of shopping mall development on small township retailers. South African Journal of Economic and Management Sciences, 11(1), 37–53.
17. Pick n Pay Holdings Ltd. (2023). Integrated Annual Report. [Online] Available at: https://www.picknpayinvestor.co.za
18. South African Government. (2021). Protection of Personal Information Act (POPIA). [Online] Available at: https://www.gov.za/documents/protection-personal-information-act
19. Naude, M.J., & Chiweshe, N. (2017). Challenges faced by SMEs in South Africa: Are HRM practices a solution? Problems and Perspectives in Management, 15(3), 215–223
20. Klein, H.K., & Myers, M.D. (1999). A set of principles for conducting and evaluating interpretive field studies in information systems. MIS Quarterly, 23(1), 67–94.
21. Yin, R.K. (2018). Case Study Research and Applications (6th ed.). Sage Publications.
22. Eisenhardt, K.M. (1989). Building theories from case study research. Academy of Management Review, 14(4), 532–550.
23. Walsham, G. (1995). Interpretive case studies in IS research: Nature and method. European Journal of Information Systems, 4(2), 74–81.
24. Bowen, G.A. (2009). Document analysis as a qualitative research method. Qualitative Research Journal, 9(2), 27–40.
25. Flick, U. (2007). Managing Quality in Qualitative Research. Sage.
26. Braun, V., & Clarke, V. (2006). Using thematic analysis in psychology. Qualitative Research in Psychology, 3(2), 77–101.
27. Miles, M.B., Huberman, A.M., & Saldaña, J. (2014). Qualitative Data Analysis: A Methods Sourcebook (3rd ed.). Sage.

28. Lincoln, Y.S., & Guba, E.G. (1985). Naturalistic Inquiry. Sage Publications.
29. Charmaz, K. (2014). Constructing Grounded Theory (2nd ed.). Sage.
30. Shenton, A.K. (2004). Strategies for ensuring trustworthiness in qualitative research projects. Education for Information, 22(2), 63–75.
31. Denzin, N.K. (1978). The Research Act: A Theoretical Introduction to Sociological Methods. McGraw-Hill.
32. Klein, H.K. and Myers, M.D. (1999). A Set of Principles for Conducting and Evaluating Interpretive Field Studies in Information Systems. MIS Quarterly, 23(1), 67-93.
33. Walsham, G. (1995). Interpretive case studies in IS research: nature and method. European Journal of Information Systems, 4(2), 74-81.
34. Greenhalgh, T., Robert, G., Macfarlane, F., Bate, P., & Kyriakidou, O. (2004). Diffusion of innovations in service organizations: systematic review and recommendations. The Milbank Quarterly, 82(4), 581-629.